\documentclass[12pt, prd, showpacs]{revtex4}
\usepackage{amssymb}
\usepackage{amsmath}

\input{tcilatex}

\begin{document}

\title{Comment on "Black Holes are Neither Particle Accelerators Nor Dark
Matter Probes"}
\author{O. B. Zaslavskii}
\affiliation{Department of Physics and Technology, Kharkov V.N. Karazin National
University, 4 Svoboda Square, Kharkov 61022, Ukraine}
\email{zaslav@ukr.net }

\begin{abstract}
The Banados-Silk-West effect consists in the possibility to get infinite
energy in the centre of mass frame of two particles colliding near the black
hole horizon. According to S. T. McWilliams, Phys. Rev. Lett. \textbf{110}
(2013) 011102, the energy at infinity of the outcome vanishes because of
infinite redshift when the point of collision approaches the horizon. I show
that this is not so.
\end{abstract}

\maketitle

The effect discovered by Ba\~{n}ados, Silk and West (BSW) states that if two
particles collide near the black hole horizon, the energy in their centre of
mass frame $E_{c.m.}$ can grow unbound \cite{ban}. For astrophysical
purposes, it is important to know what can be detected at infinity as the
products of collision. Here, there are two kinds of relevant quantities: (i)
fluxes from a vicinity of the horizon, (ii) masses $m$ and energies $%
E_{\infty }$ of particles. In both cases strong redshift is crucial. Its
account in a recent work \cite{1} lead to the conclusions that (i) fluxes
vanish due to relativistic dilatation of time, (ii) $E_{\infty }$ $%
\rightarrow 0$. Conclusion (i) looks reasonable and is essential for
estimates of expected fluxes. However, (ii) is incorrect. 

Let particles 1 and 2 collide to produce particles 3 (escapes) and 4 (falls
into a black hole). Eq. 6 of \cite{1} reads%
\begin{equation}
E_{\infty }=\alpha E_{c.m.}  \label{6}
\end{equation}
where $\alpha $ is the lapse due to geodesic motion. For a general
stationary axially symmetric black hole $\alpha =\left( u^{0}\right) ^{-1}=%
\frac{N^{2}}{E_{\infty }-\omega L}$, $\omega =-g_{0\phi }/g_{00}$, $N$ is
the lapse function entering the metric. Near the horizon$,N\rightarrow 0$, $%
E_{c.m..}\sim \frac{1}{\sqrt{N}}$ \cite{prd} but $\alpha $ oversomes it, so
eq. (\ref{6}) leads to the conclusion that $\left( E_{\infty }\right)
_{3}\rightarrow 0$ for any $L$ (not only for the case $L=0$ in eq. 7 of \cite%
{1}). Meanwhile, the correct formula is%
\begin{equation}
\left( E_{\infty }\right) _{3}-m_{3}\left( u_{i}\right) _{3}\left(
u^{i}\right) _{4}\alpha =\alpha E_{loc.}=\alpha m_{3}\gamma (3,4)\text{, }
\label{e3}
\end{equation}%
$\gamma (3,4)=-\left( u_{\mu }\right) _{3}\left( u^{\mu }\right) _{4}$ is
the Lorentz factor of relative motion, $\alpha $ refers to particle 4, $\mu
=0,i$, $E_{c.m.}^{2}=m_{3}^{2}+m_{4}^{2}+2m_{3}m_{4}\gamma (3,4)$.

It follows from the geodesic equations and (\ref{e3}) that for the
near-horizon collision,  $\left( E_{\infty }\right) _{3}\approx \omega
_{H}L_{3}$, $\omega _{H}$ is the black hole angular velocity. This is in
perfect agreement with the previous reuslts \cite{p} - \cite{z}. Thus strong
redshift is quite compatible with nonzero $\left( E_{\infty }\right) _{3}$ .
The incorrect conclusion about $\left( E_{\infty }\right) _{3}\rightarrow 0$
was based on (i) confusion between $E_{c.m.}$ and $E_{loc.}$ and (ii)
omission of kinetic terms in (\ref{e3}). Thus for individual collisions
there are no more restrictive bounds than those already obtained in \cite{p}
- \cite{z}.

\end{document}